\RequirePackage{etex}
\documentclass{iau}

\usepackage{amsmath}
\usepackage{graphicx}
\usepackage{multirow}

\newcommand{\one}{$^1$}
\newcommand{\two}{$^2$}
\newcommand{\thr}{$^3$}
\newcommand{\fou}{$^4$}
\newcommand{\fiv}{$^5$}
\newcommand{\six}{$^6$}
\newcommand{\sev}{$^7$}
\newcommand{\eig}{$^8$}

\begin{document}

\lefttitle{Zhang et al.}
\righttitle{First results from JACKS}

\jnlPage{1}{6}
\jnlDoiYr{2026}
\doival{10.1017/xxxxx}

\aopheadtitle{Proceedings IAU Symposium}
\editors{M. Zaja\v{c}ek, T. Je\v{r}\'abkov\'a, V. Karas, R. Schödel \& P. Sukov\'a, eds.}

\title{First Results from JACKS: A VLA Ammonia View of Gas Heating and Star-forming Cores in the Central Molecular Zone}

\author{Xu Zhang\one, Xing Lu\one, Elisabeth A. C. Mills\two, Qizhou Zhang\thr, Daniel L. Walker\fou, Natalie O. Butterfield\fiv, Namitha Issac\one, Annika Schmiedeke\six, Suinan Zhang$^{7,8}$\\
and JACKS collaborations}
\affiliation{\one Shanghai Astronomical Observatory, Chinese Academy of Sciences, Shanghai, China; \email{zhangxu@shao.ac.cn}\\
\two Department of Physics and Astronomy, University of Kansas, 1251 Wescoe
Hall Drive, Lawrence, KS 66045, USA\\
\thr Center for Astrophysics | Harvard \& Smithsonian, 60 Garden Street, Cambridge, MA, 02138, USA\\
\fou UK ALMA Regional Centre Node, Jodrell Bank Centre for Astrophysics,
Oxford Road, The University of Manchester, Manchester M13 9PL, UK\\
\fiv National Radio Astronomy Observatory, 520 Edgemont Road, Charlottesville, VA 22903, USA\\
\six Green Bank Observatory, 155 Observatory Road, Green Bank, WV 24944,
USA\\
\sev National Astronomical Observatory of Japan, National Institutes of Natural Sciences, Mitaka, Tokyo
181-8588, Japan\\
\eig Department of Earth and Planetary Sciences, Institute of Science Tokyo, Meguro, Tokyo 152-8551, Japan}

\begin{abstract}
The Central Molecular Zone (CMZ) combines warm molecular gas, strong turbulence, and a star-formation rate below that expected from its dense-gas reservoir. The JVLA Ammonia C-configuration K-band Survey (JACKS) is designed to connect these environmental conditions to the thermal and kinematic structure of compact, potentially star-forming gas. JACKS follows the footprint of the ALMA Central Molecular Zone Exploration Survey (ACES) with 31 Very Large Array mosaic fields.  Its K-band setup covers the metastable NH$_3$ $(1,1)$--$(5,5)$ inversion transitions, five non-metastable NH$_3$ lines, the 22-GHz H$_2$O maser transition, H64$\alpha$ and H66$\alpha$, and broad continuum windows.  The nominal resolutions are approximately $1.5''$ ($0.06$ pc at the Galactic Centre) and $0.4$ km s$^{-1}$ for the NH$_3$ lines.  We present preliminary imaging products that demonstrate water-maser self-calibration, compact and extended continuum emission, filamentary NH$_3$ structure, and complex compact sources with multiple velocity components.  Tests that combine JACKS with the SWAG and HOPS ammonia surveys recover progressively larger angular scales.  These early products establish the survey's ability to connect core-scale gas heating and kinematics to the cloud-scale CMZ environment; quantitative temperature, opacity, and core-population measurements will follow from the uniformly reduced data set.
\end{abstract}


\maketitle

\section{The CMZ and the need for a high-resolution ammonia survey}

The CMZ is a unique environment for star formation.  Molecular gas temperatures are commonly of order 50--100 K or higher, cloud-scale line widths reach roughly 10--30 km s$^{-1}$, and even compact structures show line widths of several km s$^{-1}$ \citep{ginsburg2016,krieger2017,shetty2012,henshaw2016,kauffmann2017}.  At the same time, star formation is suppressed relative to simple expectations based on the available dense gas.  Resolving how heating, turbulence, and gravitationally concentrated structure vary from clouds to cores therefore requires a tracer that is both temperature-sensitive and observable over a wide range of angular scales.

Ammonia provides a set of closely spaced inversion transitions that can constrain excitation temperature while tracing dense molecular gas.  Existing surveys supply essential large-scale context.  HOPS mapped NH$_3$ with the Mopra telescope at about $2'$ resolution \citep{purcell2012}; SWAG used the Australia Telescope Compact Array to image the CMZ ammonia ladder at approximately $20''$ resolution \citep{krieger2017}.  JACKS adds an approximately $1.5''$ view, corresponding to about 0.06 pc, while matching the ACES footprint.  ACES offers complementary 3-mm molecular-line imaging at comparable angular resolution \citep{longmore2026} and reveals the highly filamentary structure of the CMZ \citep{battersby2026}.

\section{Survey design}

JACKS comprises 31 mosaic fields spanning the ACES footprint; each field contains 14 VLA pointings. The correlator configuration includes the metastable NH$_3$ $(1,1)$ through $(5,5)$ lines, five non-metastable NH$_3$ transitions, a 22-GHz H$_2$O maser window, H64$\alpha$ and H66$\alpha$, and approximately 3 GHz of continuum bandwidth distributed between the two sidebands. The NH$_3$ channels provide a velocity resolution of approximately 0.4 km s$^{-1}$.

The survey is intended to bridge spatial scales of lower-resolution data. Interferometric JACKS images focus on compact structures, whereas SWAG and HOPS supply successively shorter spatial frequencies. Joint imaging can therefore retain the VLA resolution while restoring emission that is resolved out by the C-configuration data alone.

\section{Preliminary results}

\subsection{Calibration, continuum, and NH$_3$ emission}

\begin{figure}
    \centering
    \includegraphics[width=0.8\linewidth]{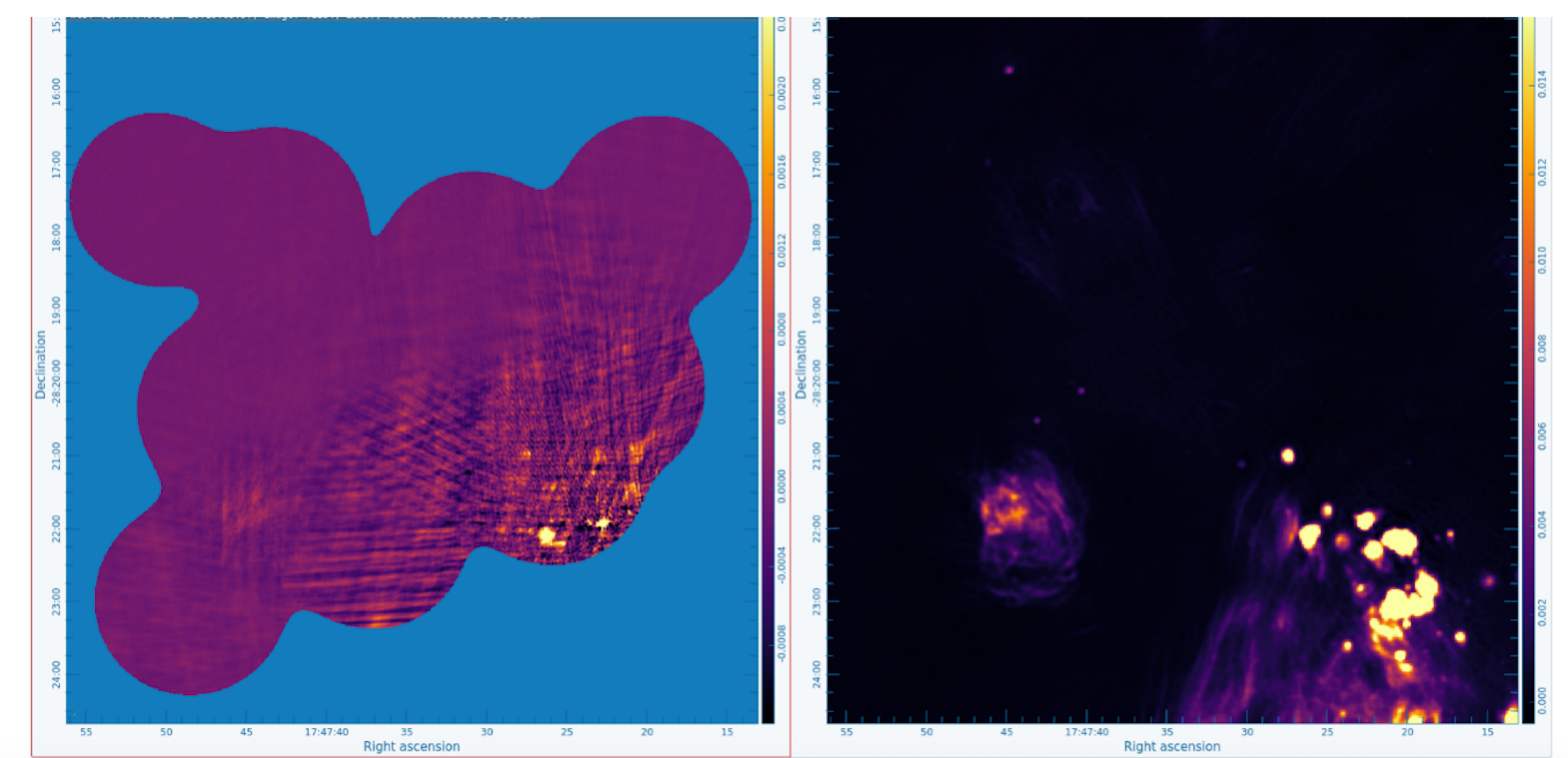}
    \caption{The radio continuum map of Field 3 (left, including Sgr B1) compared with its corresponding MeerKAT 1.3 GHz map (right).}
    \label{fig:con}
\end{figure}

Bright 22-GHz water masers are detected in the early fields and can be used for self-calibration. The resulting phase solutions improve the quality of the continuum and line products. The preliminary 1.3 cm continuum map of Sgr B1 (Field 3) is shown in the left panel of Fig. \ref{fig:con}, compared with the MeerKAT 1.3 GHz map in the right panel. A qualitative comparison with MeerKAT 1.3 GHz images highlights the complementary mixture of thermal and non-thermal structures sampled at centimeter wavelengths; a spectral classification will require matched-resolution, multi-frequency analysis and is deferred to later work.

\begin{figure}
\centering
\includegraphics[width=0.8\textwidth]{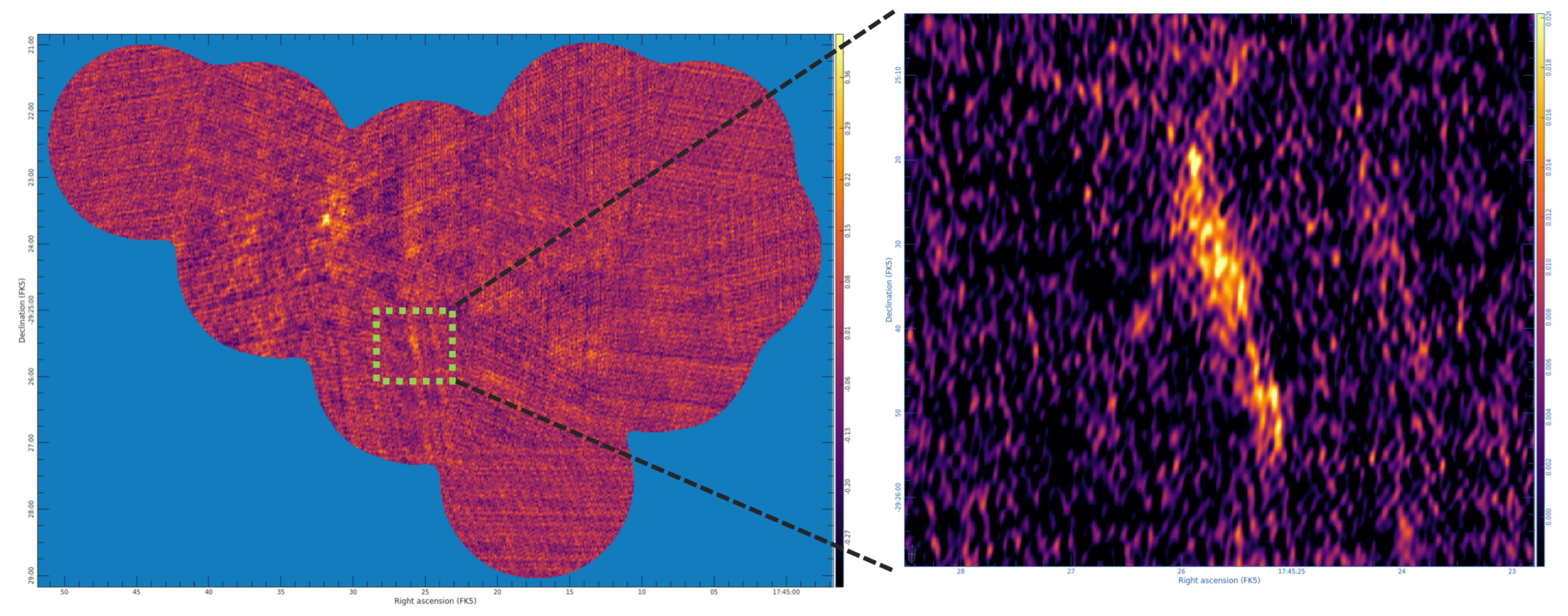}
\caption{Preliminary Field 28 NH$_3$ $(3,3)$ products.  The mosaic-scale integrated-intensity map (left) contains coherent extended emission; a zoom-in channel map near $-9.2$ km s$^{-1}$ (right) isolates a narrow filamentary feature.  Intensities and colour scales are shown as in the conference presentation; quantitative measurements await the uniform survey reduction.}
\label{fig:nh3}
\end{figure}

The left panel of Fig. \ref{fig:nh3} shows the NH$_3$ $(3,3)$ moment-0 map for Field 28. Although VLA C-configuration shows strong spatial filtering for the observation of NH$_3$, we can still find some extended faint structures in addition to the remarkable point source. The right panel is the zoom-in channel map for a filamentary structure.  

\subsection{Discovering faint structures by comparison with ACES}

\begin{figure}
\centering
\includegraphics[width=0.8\textwidth]{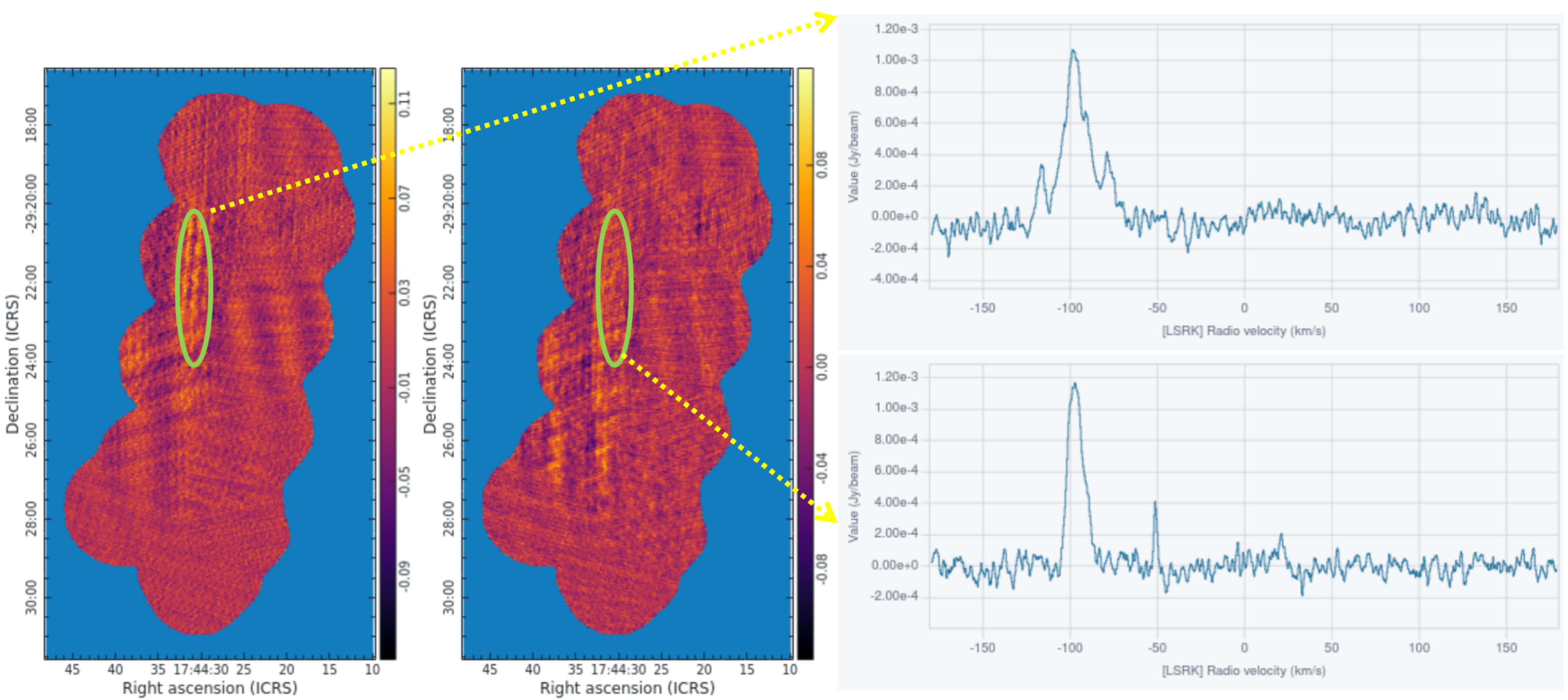}
\caption{The NH$_3$ (1,1) (left) and (3,3) (middle) moment-0 maps of Field 30. The right two panels show the integral line profiles within the green ellipses.}
\label{fig:fil}
\end{figure}

Across Field 30, NH$_3$ $(1,1)$ and $(3,3)$ emission associated with the long filament LF3 \citep{battersby2026} is only weakly detected in the integrated-intensity maps (Fig.~\ref{fig:fil}). However, spectra averaged over the LF3 footprint identified by ACES show both transitions at high signal-to-noise ratios. Their peak velocities are also consistent with the velocity of LF3 measured by \citet{battersby2026}, supporting the interpretation that the emission traces an extended, velocity-coherent dense-gas structure rather than an imaging artifact. This result demonstrates that direct comparison with the dense filament network revealed by ACES can help identify similarly faint NH$_3$ structures in other JACKS fields.

\subsection{Recovering large angular scales}

\begin{figure}
\centering
\includegraphics[width=0.9\textwidth]{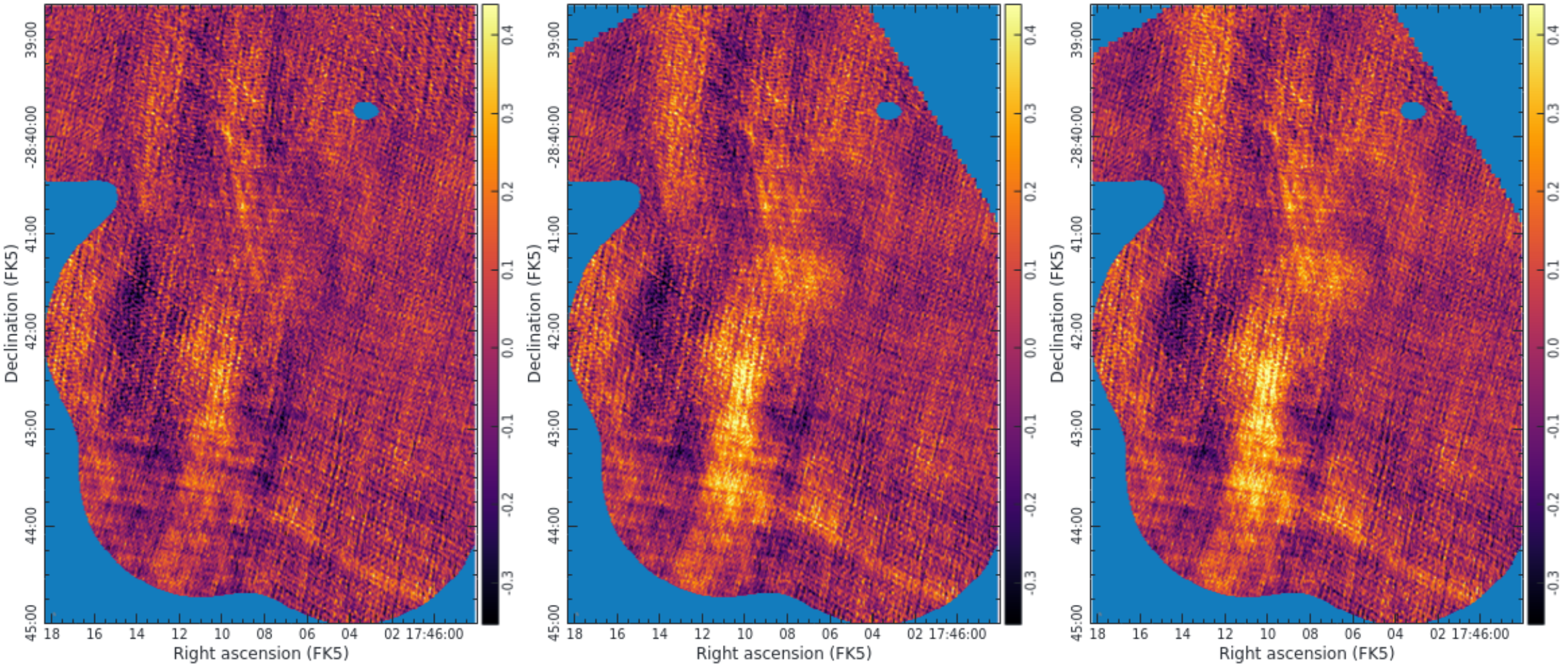}
\caption{NH$_3$ $(1,1)$ integrated-intensity maps toward the Brick: JACKS only (left), JACKS+SWAG (centre), and JACKS+SWAG+HOPS (right).  The progressively combined products recover increasingly extended emission while retaining compact structure.}
\label{fig:combine}
\end{figure}

The Brick (Field 12) provides an early demonstration of multi-survey combination.  JACKS alone emphasizes compact, high-spatial-frequency structure.  Adding SWAG restores intermediate angular scales, and including HOPS adds the broadest emission component (Fig.~\ref{fig:combine}).  The corresponding spectra show that recovered flux changes substantially when short-spacing information is introduced.  Temperature and column-density maps should therefore be derived only after the transitions have been placed on a consistent spatial-frequency footing.

\section{Outlook}

The first JACKS images validate the central survey strategy.  Maser self-calibration improves the high-resolution products; NH$_3$ inversion transitions resolve filamentary and compact structures; and combination with SWAG and HOPS recovers the cloud-scale emission required for robust excitation analysis.  The next stage is a uniform calibration and imaging pass across all 31 mosaics, followed by hyperfine and multi-transition modelling.  The primary products will include kinetic-temperature, opacity, column-density, and velocity-dispersion maps, together with catalogues of compact NH$_3$ peaks and 22-GHz water masers.  Comparison with ACES will connect these thermal and kinematic measurements to 3-mm chemistry and dense-gas morphology.  H64$\alpha$, H66$\alpha$, the non-metastable NH$_3$ transitions, and the broad-band continuum will provide additional constraints on ionized gas, radiative excitation, and embedded star formation.

\section*{Acknowledgements}

We thank the full JACKS collaboration, including Xing Lu, Elisabeth A. C. Mills (PI), Qizhou Zhang, Daniel L. Walker, Anika Schmiedeke, Natalie O. Butterfield, Namitha Issac, and collaborators. We also thank the VLA observation-support team and the organizers of the GC2026 conference.  The project homepage is \texttt{https://mills.ku.edu/JACKS}.

\end{document}